\documentclass[aps,prb,twocolumn,showpacs,amsmath,amssymb,superscriptaddress,longbibliography,balancelastpage]{revtex4-1}
\usepackage{graphicx}
\usepackage{dcolumn}
\usepackage{bm}
\usepackage{natbib}
\usepackage{color}
\usepackage[normalem]{ulem}

\begin{document}

\title{Optical evidence for a Weyl semimetal state in pyrochlore Eu$_2$Ir$_2$O$_7$} 

\author{A. B. Sushkov}
\affiliation{Center for Nanophysics and Advanced Materials, Department of Physics, University of Maryland, College Park, Maryland 20742, USA}
\author{J. B. Hofmann}
\affiliation{Condensed Matter Theory Center and Joint Quantum Institute, Department of Physics, University of Maryland, College Park, Maryland 20742, USA}
\author{G. S. Jenkins}
\affiliation{Center for Nanophysics and Advanced Materials, Department of Physics, University of Maryland, College Park, Maryland 20742, USA}
\author{J. Ishikawa}
\affiliation{Institute for Solid State Physics, University of Tokyo, Kashiwa, Chiba 277-8581, Japan}
\author{S. Nakatsuji}
\affiliation{Institute for Solid State Physics, University of Tokyo, Kashiwa, Chiba 277-8581, Japan}
\affiliation{PRESTO, Japan Science and Technology Agency (JST), 4-1-8 Honcho Kawaguchi, Saitama 332-0012, Japan}
\author{S. Das Sarma}
\affiliation{Condensed Matter Theory Center and Joint Quantum Institute, Department of Physics, University of Maryland, College Park, Maryland 20742, USA}
\author{H. D. Drew}
\affiliation{Center for Nanophysics and Advanced Materials, Department of Physics, University of Maryland, College Park, Maryland 20742, USA}

\date{\today}

\begin{abstract}
A Weyl semimetallic state with pairs of nondegenerate Dirac cones in three dimensions was recently predicted to occur in the antiferromagnetic state of the pyrochlore iridates. 
Here, we show that the THz optical conductivity and temperature dependence of the free carrier response in pyrochlore Eu$_2$Ir$_2$O$_7$ match the predictions for a Weyl semimetal and suggest novel Dirac liquid behavior.  
The interband optical conductivity vanishes continuously at low frequencies signifying a semimetal. 
The metal-semimetal transition at $T_N=110$~K is manifested in the Drude spectral weight, which is independent of temperature in the metallic phase, and which decreases smoothly in the ordered phase.
The temperature dependence of the free carrier weight below $T_N$ is in good agreement with theoretical predictions for a Dirac material.  
The data yield a Fermi velocity $v_F \approx 4 \cdot 10^7$~cm/s, a logarithmic renormalization scale $\Lambda_L \approx 600$~K, and require a Fermi temperature of $T_F \approx 100$~K associated with residual unintentional doping to account for the low temperature optical response and dc resistivity.
\end{abstract}

\maketitle

Condensed matter physics is in an exciting new era dominated by Dirac materials and topological effects.
The introduction of topology, particularly in association with spin-orbit coupling, has led to predictions of novel electronic and optical phenomena~\cite{Pesin2010Nature,Qi_Zhang_2011,Hasan2010RMP,Hasan_Moore_2011}. 
The most recent development is the discovery of three-dimensional (3D)  Dirac~\cite{Xu2013ArXiv,Liu2014Science,Neupane2014NatComm,Liu2014NatMater,Borisenko2014PRL,Xu_2015} and Weyl~\cite{Lv2015ArXiv,Xu_Science2015,Xu2015ArXiv} materials, where the Dirac cones come in pairs and have non-Kramers degenerate chiral bands protected by topology.  
A 3D Weyl state has been predicted~\cite{Wan_2011} for the $5d$ transition metal oxide family of pyrochlore iridates~\cite{Witczak-Krempa_2014}, which are strongly interacting materials with a strong spin-orbit interaction. 
However, in the absence of direct experimental evidence their ground state is still under intense discussion~\cite{Witczak-Krempa_2013,
Ishii_Mizuta_Kato_Ozaki_Weng_Onoda_2015,
Shinaoka_Hoshino_Troyer_Werner_2015,Zhang_Haule_Vanderbilt_2015}.

The pyrochlore iridates, $R_2$Ir$_2$O$_7$, are strongly interacting materials that exhibit frustrated magnetism and an associated metal-insulator transition as $R$ varies across the rare earth series~\cite{Matsuhira_2011}.  
Due to a strong spin-orbit interaction these materials have been recognized as having potentially exotic ground states including the Weyl semimetallic state in the low temperature magnetic phase. 
Also, axion insulator~\cite{Wan_2011,Witczak-Krempa_2013}, topological band insulator and Mott insulator~\cite{Pesin2010Nature}, and spin liquid states~\cite{Nakatsuji_2006,Machida_2010} have been proposed. 
 X-ray diffraction experiments show that there is no structural transition breaking the cubic symmetry in Nd, Eu, and Pr pyrochlores, and that Eu compound displays a smooth thermal contraction of the lattice parameters through the "metal-insulator" transition~\cite{Takatsu_2014}. 
Ueda et al.~\cite{Ueda_2012} reported a 50 meV optical gap in the ground state 
of polycrystalline Nd$_2$Ir$_2$O$_7$ 
which was tuned to zero by the partial substitution of Ir by Rh  and that gapless state was proposed to be a Weyl semimetal. 
By contrast, infrared studies of Bi$_2$Ir$_2$O$_7$ found a strongly metallic ground state~\cite{Lee_2013}. Therefore, at present there is no clear evidence of the predicted Weyl semimetal state in the rare earth pyrochlores. 

Here, we present evidence of a Weyl semimetal on the basis of experimental and theoretical studies of the optical response,  and we identify the electronic phase transition in Eu$_2$Ir$_2$O$_7$ at $T_N$ as a metal-semimetal transition in which the  conductance is controled by the thermal population of the Weyl cones. 
Single crystals of Eu$_2$Ir$_2$O$_7$ were grown as described elsewhere~\cite{Ishikawa}.
Our crystal has the same dc conductivity as the sample \#2 of Ref.~\cite{Ishikawa}.
Fourier transform infrared reflectivity and transmission measurements were performed on the 1.8 mm in diameter (111) face of a single crystal. 
For transmission measurements at frequencies below 100 cm$^{-1}$, the crystal was glued down to an intrinsic Si substrate using transparent Stycast epoxy and was polished down to $\approx$ 10 micron.

Figure~1 shows the measured reflectivity and transmission spectra of Eu$_2$Ir$_2$O$_7$ over a broad frequency range. 
The transmission measured at 7 K immediately suggests a semimetalic state.  
The negative slope implies an optical conductivity growing with frequency.  
An insulator (or gapped state) would give a flat transmission, and a free carrier Drude term would give a slope of opposite sign. 

 We have analyzed the reflectivity and transmission spectra in two standard ways~\cite{Suppl}: first, using a model of a sum of Lorentzian oscillators and second, using a variational dielectric function (equivalent to a Kramers-Kronig transformation)~\cite{Kuzmenko_2005}. 
In the first method, we fit the reflectivity and the transmission spectra of Eu$_2$Ir$_2$O$_7$ using a Lorentzian model, in which the complex dielectric function $\varepsilon=\varepsilon_1+i\varepsilon_2$ takes the form:
\begin{equation}
    \varepsilon(\omega)=\varepsilon_{\infty}+
    \sum_j\frac{\omega_{sj}^2}{\omega_{0j}^2-\omega^2-i\omega\gamma_j} .
\label{eq:lorentz}
\end{equation}
Here $\omega_{0j}$, $\gamma_j$, and $\omega_{sj}$ are the resonance, the damping, and the spectral weight  frequencies of the $j$th electric dipole active mode, respectively. 
For the present system, we use a Drude term, seven symmetry allowed phonons~\cite{MCCAFFREY_71}, and we find that the electronic response can be well represented using five electronic transitions, the lowest one at $\omega_{0}$=306~cm$^{-1}$ with $\gamma$=1143~cm$^{-1}$. 
In the second method, we use the results of the first method as a starting spectrum for the variational $\varepsilon(\omega)$ to account for small deviations~\cite{Suppl}. 
The complex optical conductivity is related to the dielectric function by $\sigma=\sigma_1+i\sigma_2 = \omega\varepsilon/4\pi i$.

\begin{figure}
\includegraphics[width=\columnwidth]{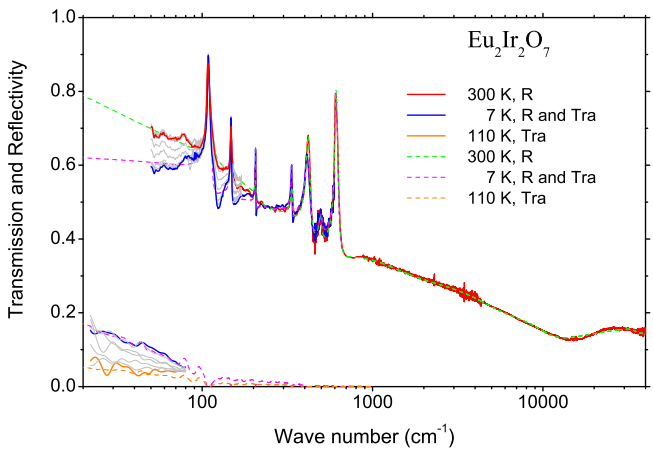}
\caption{(color online). Reflectivity (R) and transmission (Tra) of a Eu$_2$Ir$_2$O$_7$ single crystal. 
Dashed curves are fits using the Lorentzian model (Eq.~(\ref{eq:lorentz})). 
Grey lines are spectra at intermediate temperatures. 
The temperature dependence was measured over a 20--6,000 cm$^{-1}$ frequency range.} 
\label{refl}
\end{figure}   
Figure~2 presents the frequency-dependent optical conductivity $\sigma_1$ and the dielectric constant $\varepsilon_1$ as obtained from the analysis of the experimental spectra. 
\begin{figure}
\includegraphics[width=\columnwidth]{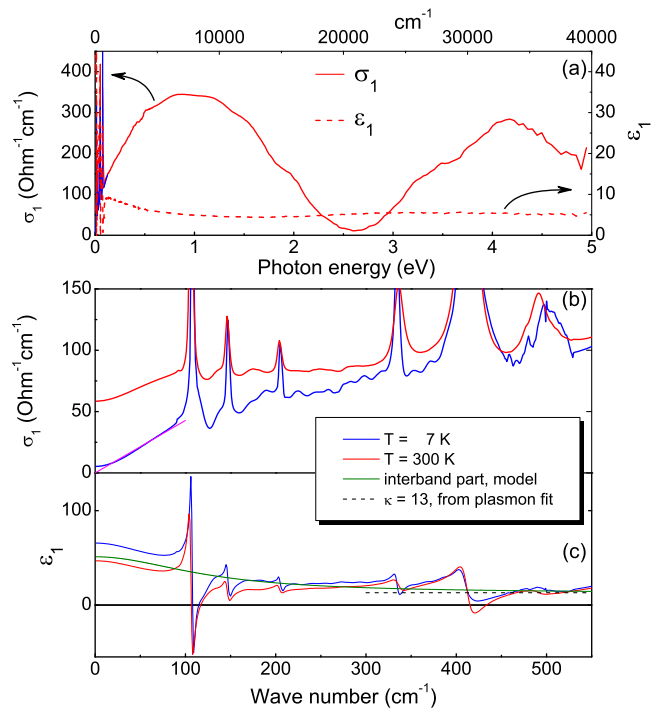}
\caption{(color online). Optical conductivity and dielectric constant of a  Eu$_2$Ir$_2$O$_7$ single crystal.  
(a) Temperature-independent broad band electronic optical conductivity. 
(b) Temperature-dependent low-frequency part of the optical conductivity.
Our zero frequency values are close to the dc measurements~\cite{Ishikawa}.
The magenta line is a linear fit of equation~(\ref{lin}) to the conductivity. 
(c) The dielectric constant. The continuous green line is the blue 7~K line with the phonon contributions removed. 
The dashed black line denotes the high-frequency dielectric constant $\kappa$ obtained from the plasmon fit.
} 
\label{fig:conductivity}
\end{figure}
Figure~2(a) shows the optical conductivity in a broad frequency range as obtained from the variational $\varepsilon$ analysis. 
The two-peak conductivity spectra are consistent with resonant inelastic X-ray scattering spectra~\cite{Hozoi_2014} and band structure calculations~\cite{Witczak-Krempa_2013}.
The vanishing of $\sigma_1$ at low frequencies is a signature of a Dirac semimetal.
The 1~eV band consists of transitions within the $J_{eff}$=1/2 bands~\cite{Witczak-Krempa_2013}. 
Figures 2(b) and 2(c) show the optical conductivity and the dielectric constant at low frequencies. 
The increasing $\varepsilon_1$ at low frequencies (Fig.~2(c)) is another  characteristic of a semimetal and is expected to diverge logarithmically for an intrinsic Weyl semimetal in the absence of disorder~\cite{throckmorton15}.  
The optical conductivity of an intrinsic Weyl semimetal can be written as
\begin{equation} 
\sigma_1= \frac{g}{12} \cdot \frac{e^2}{h} \cdot \frac{\omega}{\bar{v}_F}
\label{lin} 
\end{equation}
arising from interband excitations within the Dirac cone~\cite{Wan_2011}. 
Here, we denote by $g$ the degeneracy of Dirac nodes and by $\bar{v}_F = (v_F^x v_F^y v_F^z)^{1/3}$ the geometric mean of the Fermi velocities.  
By fitting this expression to the optical conductivity (magenta line in Fig.~2(b)) using $g$=24~\cite{Wan_2011}, 
we obtain $\bar{v}_F = 3.4\cdot10^7$~cm/s.

Based on this fit to the optical conductivity, we can extract the Coulomb interaction strength in the semimetal as given by the effective Dirac fine structure constant $\alpha$=$e^2/\hbar v_F \kappa$. 
Here, $\kappa$ which is of the order of 10 is the high-frequency dielectric constant which shows no temperature-dependence as is evident in Fig.~2(c). 
(We use $\epsilon$ and $\kappa$ to designate experimental and theoretical dielectric constant, respectively.)
This gives a bare value of $\alpha$=0.7 for the interaction strength, indicating that electron interaction could have considerable impact on the low-energy properties of this Dirac material. 
Since the corresponding fine structure constant for quantum electrodynamics is 1/137, the current problem is a strong-coupling (and non-relativistic) version of quantum
electrodynamics.

The conductivity exhibits a temperature dependence only at low frequencies $\omega$$<$500~cm$^{-1}$. 
Fits of the Eq.~(\ref{eq:lorentz}) to the experimental spectra reveal that only the Drude and phonon terms are temperature dependent while the interband conductivity can be kept constant at all temperatures. 
The validity of this conclusion is verified by a spectral weight analysis presented in the Supplemental Material~\cite{Suppl}.
We show the result for the experimental Drude spectral weight and scattering $\gamma$ as a function of temperature in Fig.~3(a,b). 
The metal-semimetal transition at $T_N$ is manifested as a smooth decrease of the Drude spectral weight in the Weyl state which can exist only below $T_N$ where time- reversal symmetry is broken~\cite{Wan_2011}.

In Fig.~3(a), we compare the experimental Drude weight with the theoretical prediction for a Dirac liquid. 
The low-temperature saturation indicates a finite Fermi energy $E_F$ (associated with unintentional residual doping), leads to a low temperature dc conductivity and marks a transition between an extrinsic metallic low-temperature regime ($k_B T<E_F$) and an intrinsic semimetallic high-temperature regime ($k_B T>E_F$). 
 Our observations are consistent with the Drude weight as predicted by kinetic theory
\begin{equation}
\omega_{sD}^2= - \frac{e^2 \bar{v}_F^2}{3} \int_{-\infty}^\infty dE \, D(E) \, \frac{\partial f(E)}{\partial E} , 
\label{eq:drudekinetic}
\end{equation}
where $D(E) = g E^2/[2 \pi^2 (\hbar \bar{v}_F)^3]$ is the density of states and $f(\varepsilon)$ the Fermi-Dirac distribution. A fit of the Drude weight (solid black line in Fig.~3(a)) gives a finite Fermi energy of 
$E_F/k_B$$\approx$70~K
 and a Fermi velocity of $\bar{v}_F$=4$\cdot$10$^7$~cm/s, in agreement with the direct fit to the frequency-dependent conductivity.  
For comparison, we include a fit to the high-temperature Drude weight in the strictly intrinsic limit (dashed blue line in Fig.~3(a)) which shows that finite doping does not affect the Drude weight at high temperature ($k_BT>E_F$). 
Note that for a general semimetallic dispersion $E \sim |k|^z$, the Drude weight is expected to scale as $\omega_{sD}^2 \sim T^{1+1/z}$ with temperature~\cite{DasSarmaPRB2015}. 
The superlinear temperature dependence implies $z<1$ in this effective model, compellingly excluding, for example, a parabolic semimetallic ground state.

Our experimental Drude weight is seen to deviate from the strict linear temperature scaling of a noninteracting semimetal, which can be attributed to electron interactions: the interaction strength of an intrinsic Dirac semimetals acquires a scale-dependence due to ultraviolet renormalization, giving rise to a superlinear temperature dependence of the Drude weight and providing a direct signature of electron interaction effects~\cite{Hofmann_Sarma_2015,throckmorton15}. 
Within RPA, the temperature dependence of the interaction strength is given by~\cite{Hofmann_Sarma_2015,Kharzeev_2014}
\begin{equation}
\alpha(T) = \frac{3\pi}{g} \ln^{-1} \frac{\Lambda_L}{T} . 
\label{eq:chargeren}
\end{equation}
Instead of depending on $e^2$, $\bar{v}_F$, and $\kappa$ separately, this expression contains a single renormalized energy scale $\Lambda_L$, the Landau pole, at which the coupling $\alpha$ diverges. 
This Landau scale is an effective parameter of the system that can be extracted from measurements of the Drude weight  
and corresponds to a cutoff scale beyond which the Dirac dispersion is no longer linear. 
In the present system, it can be associated with the Lifshitz saddle point between each pair of Weyl points~\cite{Hofmann_Sarma_2015}. 

\begin{figure}
\includegraphics[width=\columnwidth]{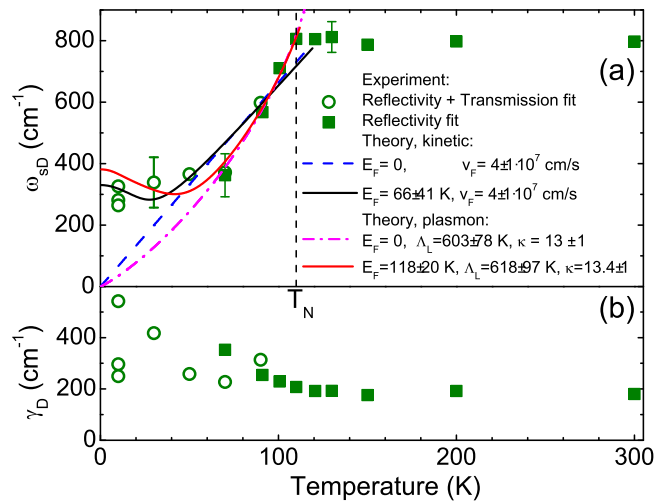}
\caption{(color online). Temperature dependence of the spectral Drude weight and scattering rate $\gamma$.
(a) Open circles --- results of simultaneous fit of reflectivity and transmission, solid squares --- results of reflectivity fits, both using Eq.~(\ref{eq:lorentz}), lines --- fits of kinetic theory (Eq.~(\ref{eq:drudekinetic})) and plasmon to these experimental points below $T_N$. 
(b) Experimental scattering rate of free carriers. 
} 
\label{wp}
\end{figure}
In the absence of disorder and phonon contributions, the Drude weight corresponds to the plasmon frequency as defined by the zero of the dielectric function $\varepsilon(\omega) = \kappa_0 - \omega_{sD}^2/\omega^2$. 
We assume $\kappa_0$ independent of $T$ and $\omega$ (Fig.~2(c)) and it is a good assumption in the frequency range 300--800~cm$^{-1}$ where solutions for $\omega_{sD}$ are found. 
We perform a fit to the experimental $\omega_{sD}$ using the plasmon dispersion of an extrinsic Dirac semimetal~\cite{Hofmann_Sarma_2015}, which fully takes into account the competing effects of finite doping and renormalization of the interaction. 
Since theoretical calculations only predict the renormalization of the effective fine structure constant $\alpha$, there is a question whether renormalization effects should be manifested in the Drude weight, the dielectric constant, or both. 
The result of the fit is shown in Fig.~3(a) as a red solid line. 
We obtain good agreement with the experiment with an effective static dielectric constant of $\kappa_0$$\approx$13 and a Fermi energy of $E_F/k_B$$\approx$100~K, which is consistent with the experimental dielectric function $\varepsilon_1$ (Fig.~2), and the fit to the Drude-Boltzmann weight (Eq.~\ref{eq:drudekinetic}), respectively. 
The Landau pole takes an anomalously small value of 
$\Lambda_L$$\approx$600~K, giving rise to strong superlinear scaling. 
This could be taken as an indirect sign that the material hosts pairs of Weyl cones in close proximity, consistent with the picture of a Weyl semimetal state with broken time-reversal symmetry.
For comparison, we also include a fit of the intrinsic finite-temperature plasmon in Fig.~3(a) (dash-dotted magenta line). 
As before, the intrinsic result captures the high-temperature behavior where the Weyl state is most expected and the saturation at low temperature can be attributed to finite doping.
Intuitively, one could expect a characteristic step in $\sigma_1(\omega)$ at 2$E_F$$\approx$140~cm$^{-1}$ but it should be smeared out by large scattering rate and cannot be seen clearly in Fig.~2(b).

Linear (or quasi-linear) dependence of the low frequency optical conductivity --- a signature of a Dirac semimetallic state --- has also been recently reported in several other 3D materials: HgCdTe~\cite{Orlita_2014}, quasicrystals~\cite{Timusk_2013}, and ZrTe$_5$~\cite{Chen_2015}.  
A notable result of our work compared to previous studies is the excellent agreement of the experimental and theoretical temperature dependent Drude spectral weight providing strong evidence of a Weyl state with finite doping. 
However, the independence of the Drude parameters on temperature in the paramagnetic state above $T_N$ is surprising and merits further study.

In summary, the optical response of a Eu$_2$Ir$_2$O$_7$ single crystal reveals a semimetallic electronic structure with approximately linear frequency dependence of the optical conductivity down to 3~meV at low temperature. 
Below $T_N$, the Drude spectral weight diminishes consistent with the reduced thermal excitations of a Weyl semimetal. 
This means that the long-thought ``metal-insulator'' transition in pyrochlore iridates may actually be a metal-semimetal transition, at least, in the case of Eu$_2$Ir$_2$O$_7$.
These two data sets can be modeled assuming a Weyl state, as 24 Weyl cones with an average Fermi velocity $\bar{v}_F$=4$\cdot$10$^7$~cm/s. 
The theoretical analyses of our optical data point toward signatures of the ultraviolet renormalization expected for an interacting Dirac liquid manifesting in the super-linear temperature dependence of the Drude weight.

\acknowledgments
  This work was supported by DOE under grant No. ER 46741-–SC0005436 (G.S.J. and H.D.D.) and LPS-MPO-CMTC (J.H. and S.D.S) and by NSF grant No. DMR-1104343 (A.B.S.). It was also supported in part by PRESTO from the Japan Science and Technology Agency, by Grants-in-Aid for Scientific Research (No. 25707030) and Program for Advancing Strategic International Networks to Accelerate the Circulation of Talented Researchers (No. R2604) from the Japanese Society for the Promotion of Science. 
Some part of the work was performed at the Aspen Center for Physics, partially funded by NSF Grant No. 1066293.   
We acknowledge useful conversations with A.~Vishwanath, Y.B.~Kim, P.~Goswami, B.~Roy, D.~Maslov, N.~Perkins and B.I.~Shklovskii.
 
%

\end{document}


\title{SUPPLEMENTAL MATERIAL \\ 
Optical evidence for a Weyl semimetal state in pyrochlore Eu$_2$Ir$_2$O$_7$} 

\author{A. B. Sushkov}
\affiliation{Center for Nanophysics and Advanced Materials, Department of Physics, University of Maryland, College Park, Maryland 20742, USA}
\author{J. B. Hofmann}
\affiliation{Condensed Matter Theory Center and Joint Quantum Institute, Department of Physics, University of Maryland, College Park, Maryland 20742, USA}
\author{G. S. Jenkins}
\affiliation{Center for Nanophysics and Advanced Materials, Department of Physics, University of Maryland, College Park, Maryland 20742, USA}
\author{J. Ishikawa}
\affiliation{Institute for Solid State Physics, University of Tokyo, Kashiwa, Chiba 277-8581, Japan}
\author{S. Nakatsuji}
\affiliation{Institute for Solid State Physics, University of Tokyo, Kashiwa, Chiba 277-8581, Japan}
\author{S. Das Sarma}
\affiliation{Condensed Matter Theory Center and Joint Quantum Institute, Department of Physics, University of Maryland, College Park, Maryland 20742, USA}
\author{H. D. Drew}
\affiliation{Center for Nanophysics and Advanced Materials, Department of Physics, University of Maryland, College Park, Maryland 20742, USA}

\date{\today}

\maketitle

\section{Model of sum of Lorentzians}

We begin our analysis with fitting the optical data with a model dielectric function of a sum of Lorentzians. 
By fitting experimental reflection and transmission spectra with 
this model, we can get separately temperature dependence of the spectral weight and scattering rate of the free carriers (Drude term), as well as resonance frequencies, spectral weight, and scattering rate for phonons and electronic transitions.  
For the present system, we use a Drude term, seven expected phonons, and five electronic transitions. 
Five Lorentzian oscillators were found sufficient to give a very good representation of the electronic response.
Figure~S1(a) shows the measured reflectivity and transmission spectra of Eu$_2$Ir$_2$O$_7$ in the frequency range of phonons together with the model curves. 
These are the same spectra as in figure 1 of the paper but with linear frequency scale. 

Figure~S1(b) shows each Lorentzian term separately except for the phonons which are shown as one spectrum. 
The sum of these terms gives the net optical conductivity. 
It is clear that these components of the optical conductivity have different resonance frequencies and,especially, widths and that they do not interfere with each other in the fitting procedure. 
We note that the Drude term at 7 K (blue curve) is broad and it gives a small but finite dc conductivity. 
The fact that fact that Drude term (red curve) and the lowest electronic peak (magenta curve) have opposite slope allows the clean separation of  their contributions by fitting the transmission spectra.
\begin{figure}
\includegraphics[width=12 cm]{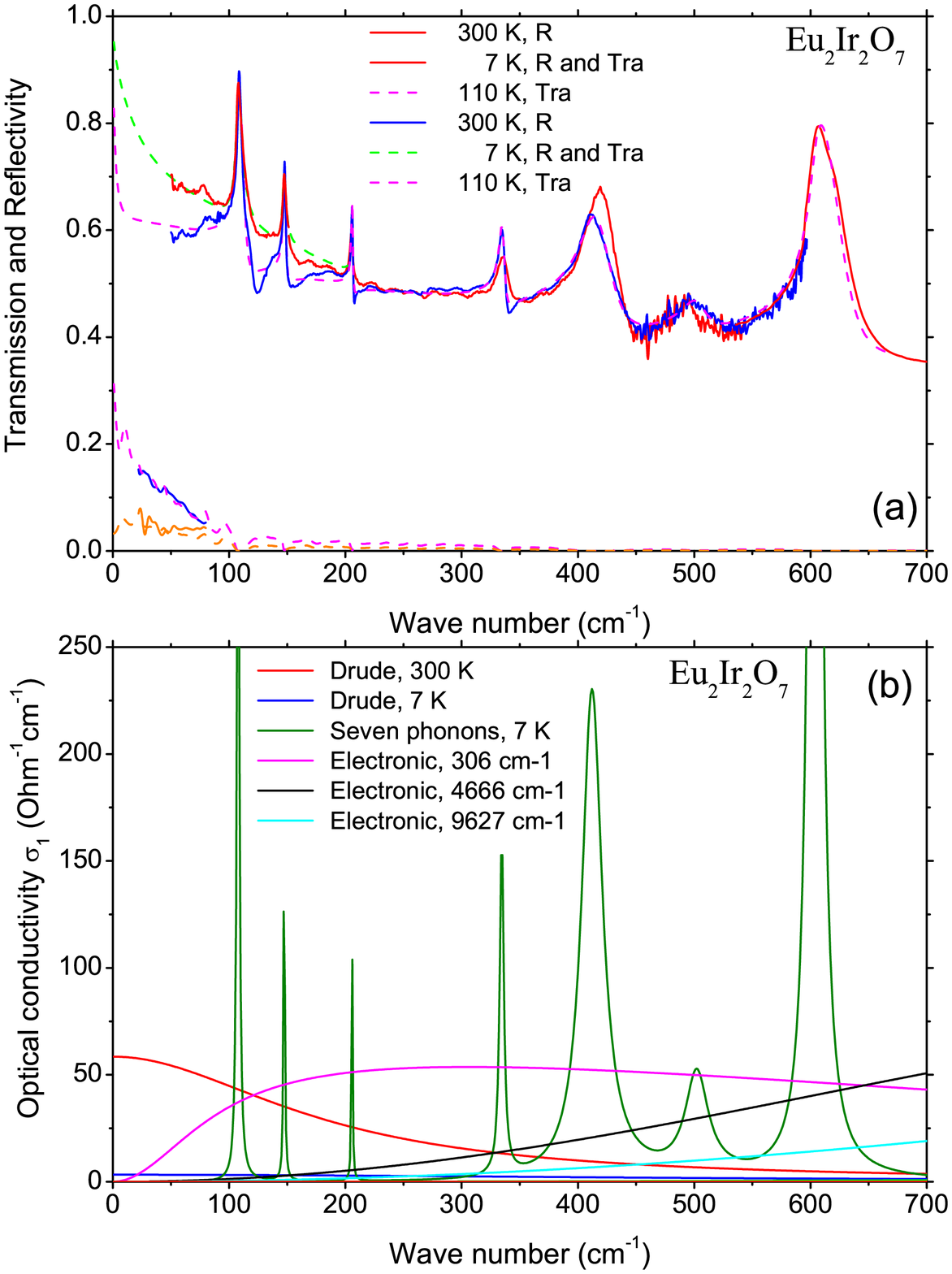}
\caption{(color online). (a) Reflectivity (R) and transmission (Tra) of a Eu$_2$Ir$_2$O$_7$ single crystal. 
Dashed curves are fits using the Lorentzian model (Eq.~(1) of the paper).  
(b) Components of the total optical conductivity: Drude, phonons, and three lowest in frequency electronic transitions.} 
\end{figure}   
We first fit simultaneously the 7 K transmission and reflection spectra. 
At this temperature, the transmission is maximum at low frequencies and it is a clear signature of semimetallic optical conductivity. 
We found that allowing for a temperature dependence of the Drude and phonon parameters is sufficient to account for the observed temperature dependence. 
These fit results are shown in Fig.~3 of the paper as open circles. 
We then fit the reflectivity spectra only at temperatures at 70 K and higher where there is no transmission for the thinned crstal at these temperatures. 
Again, a variation of only the Drude and phonon parameters keeping all other parameters fixed is sufficient. 
Results of this fit are shown in Fig.~3 as solid squares.

\section{Model of the variational dielectric function (VDF)}

In the second method, we use the results of the first method as a starting spectrum and calculate the VDF $\varepsilon(\omega)$~[27].
This accounts for small deviations between Loretzian fit and experimental data. 
Figure~S2 shows an example of the two types of fit. 
It is apparent from the figure that the model curve (blue) for VDF fit is practically indistinguishable from the experimental curve (red). 
The VDF fit is equivalent to the standard Kramers-Kronig transformation technique~[27] and both rely on assumed extrapolation beyond the measured interval. 
The value of the VDF fit is that it reproduces all small features of the spectrum not included in the sum of Lorentzians. 
\begin{figure}
\includegraphics[width=12 cm]{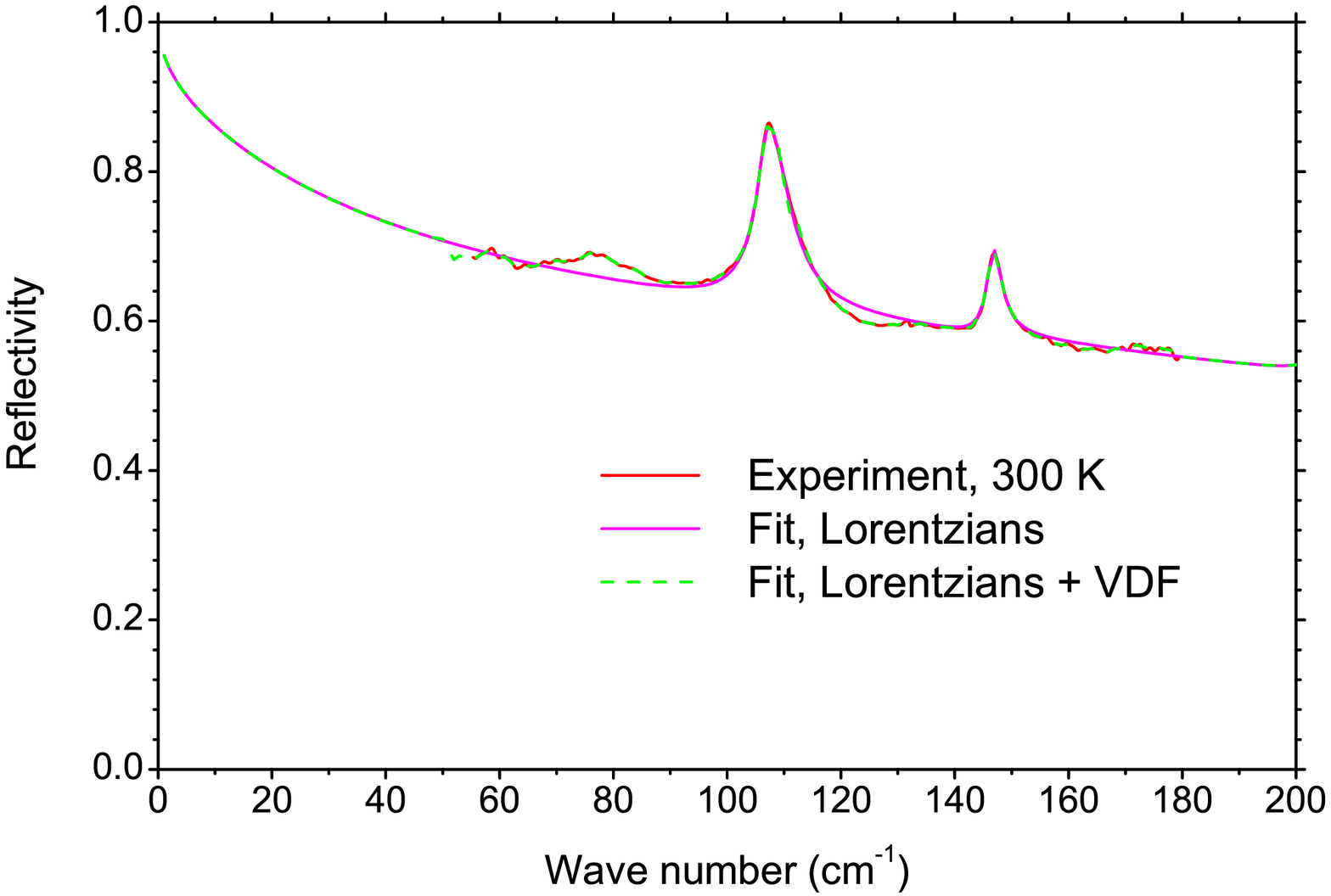}
\caption{(color online). Experimental and model reflectivity of a Eu$_2$Ir$_2$O$_7$ single crystal calculated from the full variational dielectric function.  
} 
\end{figure}   

\section{Spectral Weight}

To check our conclusion that all temperature changes come only from the Drude and phonon terms, we use another standard spectroscopic tool --- the analysis of the total spectral weight $SW(\omega)$: 
\begin{equation}
   SW(\omega_c) = \int_0^{\omega_c} \sigma_1(\omega)d\omega = \frac{\pi}{120} \omega_{pt}^2,
\end{equation}
where $\sigma_1$ is the optical conductivity in units of (Ohm$\cdot$cm)$^{-1}$ and the effective (or total) plasma frequency $\omega_{pt}$ includes all possible excitations. 
In Figure~S3, we compare $\omega_{pt}^2$ obtained by integration of the VDF optical conductivity up to the cutoff frequency $\omega_c$=530~cm$^{-1}$ with the Drude weight, $\omega_{sD}^2$ where the $\omega_{sD}$ data are the same as in Fig.~3(a) of the paper obtained from the sum of Lorentzians fit.
Here, we have shifted the $\omega_{pt}^2$ curve (blue stars) by an arbitrary constant to exclude all temperature independent excitations. 
We find that the agreement of all data is quite impressive considering that not only are they calculated differently but also they are based on different measured spectra: blue stars were obtained from the reflectivity measured with the 4 K bolometer in the frequency range 90--600~cm$^{-1}$, open circles and solid squares come from the transmission and reflection measured with the 1.6 K bolometer in the frequency range 10--200~cm$^{-1}$. 
From a comparison of the $\omega_{pt}^2(T)$ (total spectral weight) and the $\omega_{sD}^2(T)$ (Drude term), we conclude that only the Drude spectral weight depends on temperature. 
While the phonons do have temperature dependent widths their total spectral weight is practically independent of temperature compared to the Drude weight.  

\begin{figure}
\includegraphics[width=12 cm]{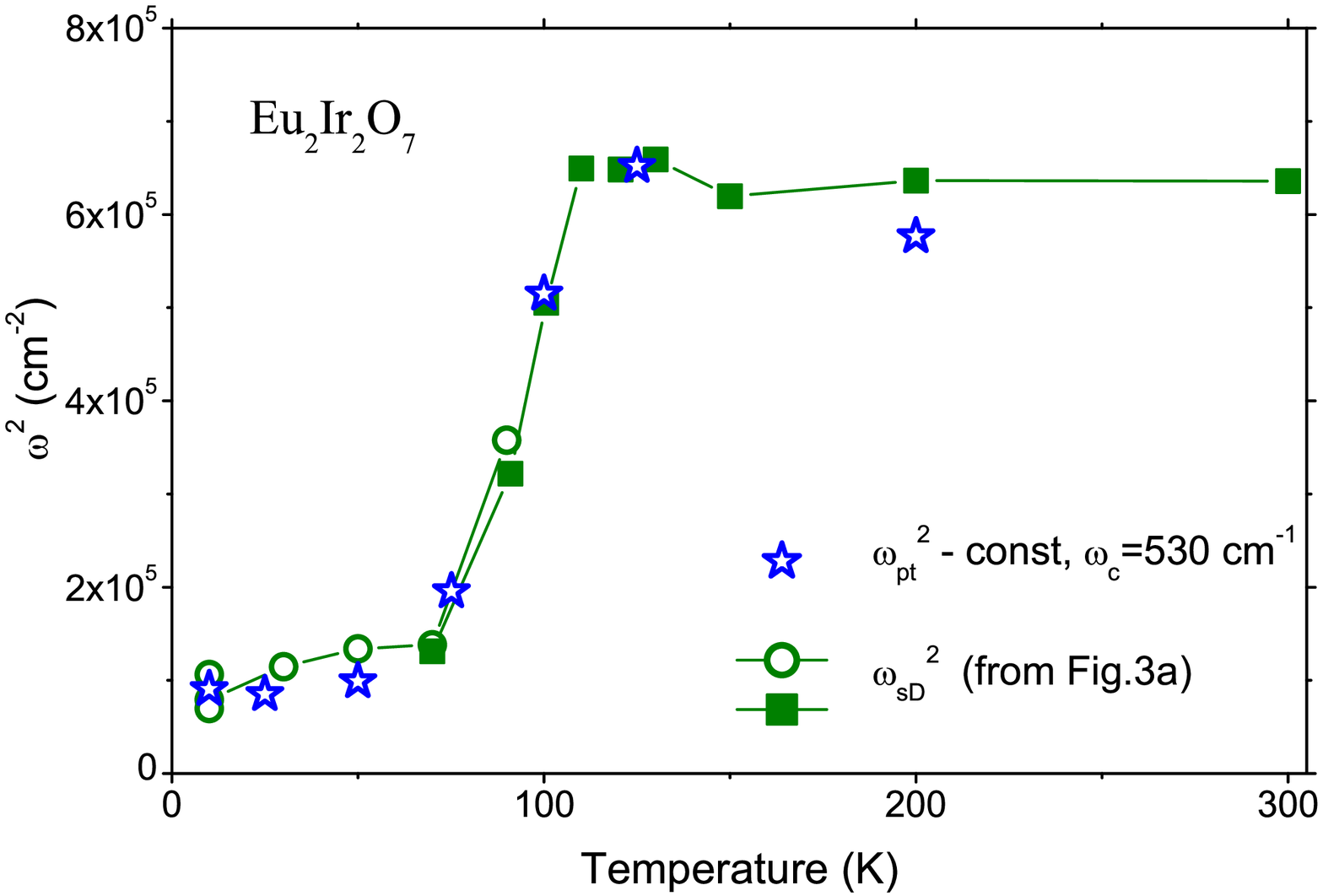}
\caption{(color online). Comparison of the temperature dependences of the total (stars) and Drude (open circles and solid squares) spectral weights. The total spectral weight was shifted down to exclude the temperature independent parts. 
} 
\end{figure}